\documentclass[twocolumn, aps,graphicx,prb, showpacs]{revtex4}
\usepackage{graphicx}
\usepackage{amssymb}
\usepackage{amsfonts}
\usepackage{amsmath}
\usepackage{color}
\usepackage{ulem}

\begin{document}

\title{Imaging of trapped ions with a microfabricated optic for quantum information processing}

\author{Erik W. Streed${^{1,\star}}$, Benjamin G. Norton$^1$, Andreas Jechow$^1$, Till J. Weinhold$^{1,2}$ \& David Kielpinski${^1}$\\
$^1$\textit{Centre for Quantum Dynamics,
Griffith University, Brisbane 4111, QLD, Australia.}\\
$^2$\textit{Centre for Quantum Computer Technology, Department of Physics,
University of Queensland, Brisbane 4072, QLD, Australia.}\\
\textit{$^\star$email: e.streed@griffith.edu.au}}

\date{\today}

% pacs 37.10.Ty- Ion Trapping
% pacs 03.67.-a Quantum information
% pracs 42.25.Fx, Diffraction and scattering

\begin{abstract}
Trapped ions are a leading system for realizing quantum information processing (QIP). Most of the technologies required for implementing large-scale trapped-ion QIP have been demonstrated, with one key exception: a massively parallel ion-photon interconnect. Arrays of microfabricated phase Fresnel lenses (PFL) are a promising interconnect solution that is readily integrated with ion trap arrays for large-scale QIP. Here we show the first imaging of trapped ions with a microfabricated in-vacuum PFL, demonstrating performance suitable for scalable QIP. A single ion fluorescence collection efficiency of $4.2\pm1.5\%$ was observed, in agreement with the previously measured optical performance of the PFL. The contrast ratio between the ion signal and the background scatter was $23\pm4$. The depth of focus for the imaging system was $19.4\pm2.4\:\mu$m and the field of view was $140\pm20\:\mu$m. Our approach also provides an integrated solution for high-efficiency optical coupling in neutral atom and solid state QIP architectures.
\end{abstract}

\maketitle

Quantum computation\cite{Shor-94,Grover-97} and communication\cite{Bennett-84} offer revolutionary solutions to challenging problems in information technology. Trapped ions are a leading system for demonstrating quantum information processing (QIP), with all basic operations\cite{Kielpinski-03, Myerson-08, Hanneke-09, Olmschenk-10} demonstrating excellent performance and a clear roadmap\cite{Kielpinski-02,Home-09} to large-scale implementations. Recent experiments\cite{Amini-10, Knoernschild-10} have demonstrated most of the technologies required by the large-scale roadmap. A key exception to this is a high efficiency ion-photon optical interconnect compatible with scaling to a massively parallel architecture. As we have previously proposed, microfabricated arrays of phase Fresnel lenses (PFLs) satisfy these criteria and are a promising solution to this problem\cite{Streed-09}.

Optical interactions are crucial to trapped-ion QIP, driving initialisation, high-speed gate operations, readout, and remote communications. A wide variety of approaches are being pursued to efficiently couple between light and individual trapped ions. These include conventional bulk optics\cite{Sondermann-07,Maiwald-09, Gerber-09, Shu-10,Piro-10,Olmschenk-10}, high finesse cavities\cite{Guthohrlein-01, Mundt-02,Leibrandt-09}, microfabricated mirrors\cite{Noek-10}, and multimode optical fibers\cite{VanDevender-10}. Highly parallel efficient coupling with conventional optics or high finesse cavities is challenging because of fabrication and alignment tolerances. Micromirrors and multimode fibers can efficiently collect sufficient light but suffer from poor single-mode coupling. In contrast, phase Fresnel lenses provide diffraction-limited high-NA coupling and can be microfabricated in large arrays on a single surface \cite{Menon-06}. Fig. 1a illustrates the proposed integration\cite{Streed-09} of PFL arrays with a scalable trapped-ion QIP architecture\cite{Kielpinski-02}. Such arrays have been used for nanolithography \cite{Menon-06} to obtain diffraction-limited performance at 28\% solid angle coverage (NA= 0.9). While PFLs, being diffractive optics, have sub-unit efficiency, diffraction efficiencies of 60\%-80\% at high NA are achievable with minimal additional fabrication complexity \cite{Cruz-Cabrera-07}.

\begin{figure*}
\includegraphics*[width=1.36\columnwidth]{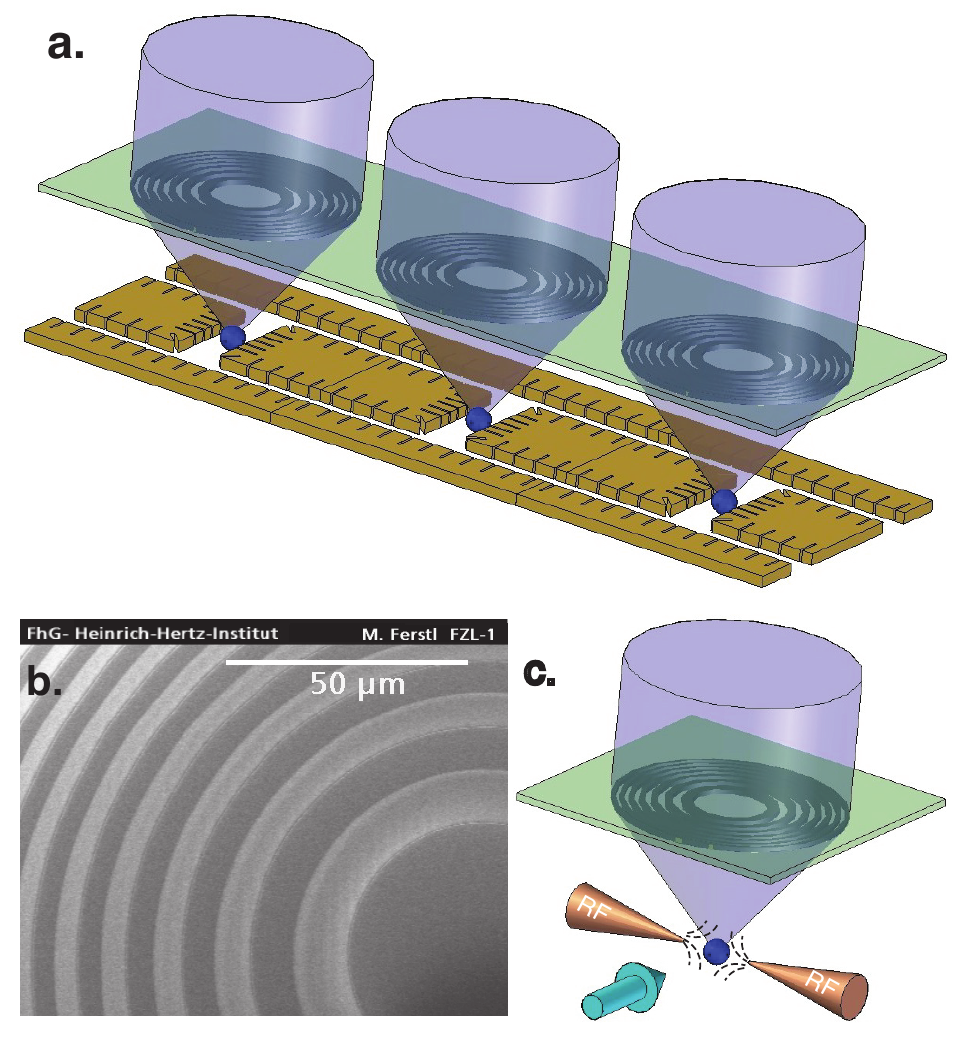}
\caption{\textbf{PFL based optical interconnects for trapped-ion QIP.} \textbf{a.} Proposed highly parallel readout of trapped-ion qubits with PFLs\cite{Streed-09}. Fluorescence from ions trapped at many sites on a microfabricated trap\cite{Kielpinski-02, Amini-10} is efficiently coupled into single optical modes by an array of microfabricated PFLs on a single substrate. \textbf{b.} Electron microscope image near the center of the PFL. The  390 nm groove depth induces $\pi$ phase shifts at $\lambda=369.5$ nm. \textbf{c.} Diagram of the experimental apparatus. A single $^{174}$Yb$^+$ ion is trapped in the RF quadrupole field (dashed lines) produced between two tungsten needles and illuminated with resonant light at $\lambda=369.5$ nm (arrow). Light scattered from the ion is collimated with an in-vacuum PFL and imaged onto a cooled CCD camera (not shown).}
\label{FigTrapAndImage}
\end{figure*}

We demonstrate the principal step towards integrating PFL arrays with ion traps through the imaging of trapped Yb$^+$ ions using a single PFL. A two-level (binary) PFL (Fig. 1b) was integrated in ultra-high vacuum with a simple yet highly flexible radiofrequency (RF) ion trap. $\mbox{Yb}^+$ ions were trapped in an RF electric quadrupole field formed by applying a potential $V_0 \cos(\Omega_{RF}t)$ with $V_0=285$ V, $\Omega_{RF}/2\pi = 20$ MHz between two tungsten needles\cite{Schrama-93, Deslauries-06} (Fig. 1c) spaced  200$\mu$m apart. Each needle was attached to a flexible welded bellows with a ceramic insulating mount to allow independent movement. Nano-positioning translation stages located outside the vacuum chamber were used to control the positioning of the needles in all three dimensions.

Loading of $^{174}$Yb$^+$ ions into the trap was performed through isotope-selective excitation of a neutral ytterbium beam with resonant 399nm light from an UV external cavity laser diode (ECLD) and subsequent photoionization with non-resonant 369.5 nm laser light. The $^{174}$Yb$^+$ ions were laser cooled on the 369.5 nm S$_{1/2}$ to P$_{1/2}$ transition using light from an ECLD\cite{Kielpinski-06} frequency stabilised to $\mbox{Yb}^+$ ions generated in an electrical discharge \cite{Streed-08}. To prevent interruption of the laser cooling, ions in the P$_{1/2}$ state which decayed into the metastable dark $D_{3/2}$ state (0.5\% branching ratio) were repumped back to the S$_{1/2}$ state by driving the transition at 935.2 nm with another ECLD. Light from all three diode lasers in the setup was delivered using single mode fibers and focused through the trap in a direction perpendicular to both the imaging and needle axes (Fig. 1c). At the center of the trap the $1/e^2$ diameter of the cooling laser was 80 $\mu$m.

The RF needles were positioned such that the PFL collimated the light scattered from the trapped ions. The ion light was then reimaged with 10$\times$ magnification onto an Andor Model DV437-BU2 cooled CCD camera. The PFL optic was microfabricated by electron-beam lithography of a fused silica substrate. A series of concentric rings, 390 nm deep, were etched into the fused silica surface to generate $\pi$ phase shifts. The resulting phase profile approximates that from a point source 3 mm from the lens with a wavelength of $\lambda=369.5$ nm. The 3 mm focal length of the lens is identical to its working distance. The pattern was written over a diameter of 5 mm, corresponding to 12\% of the total solid angle (NA=0.64). Independent profiling of the lens optical characteristics\cite{Streed-09} demonstrated a diffraction-limited subwavelength beam waist of $350\pm15$ nm ($1/e^2$ radius).  While multiple ions were observed, single ions exhibited superior lifetimes and linewidths. A magnetic field of 4 Gauss was applied along the optical axis of the PFL, to ensure that the linearly polarized 369.5 nm cooling laser traveling perpendicular to the imaging axis excited $\pi$ polarized transitions.  The background pressure in the vacuum chamber was $4\times10^{-10}$ mbar.

\begin{figure*}
\includegraphics*[width=1.3\columnwidth]{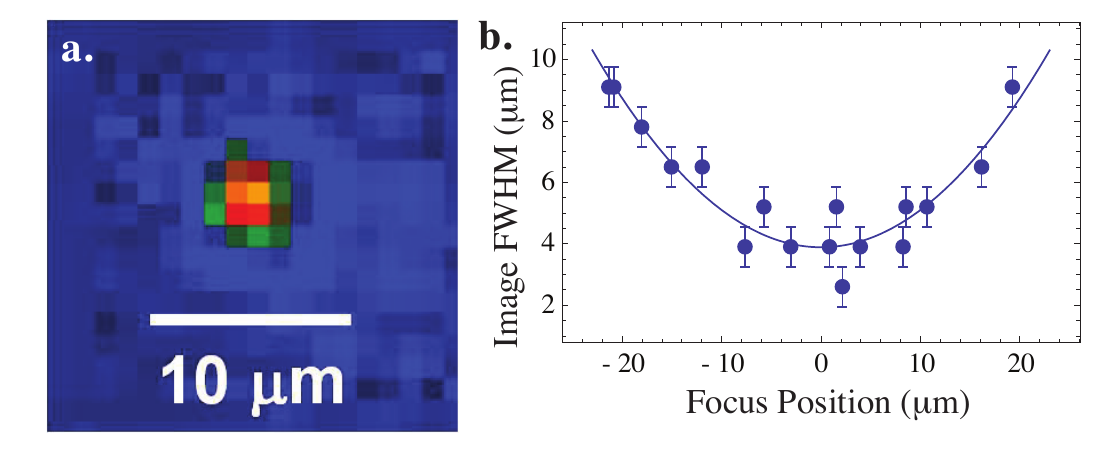}
\caption{\textbf{Imaging performance of the PFL in our  apparatus.} \textbf{a.}  Image of a single trapped $^{174}$Yb$^+$ ion using the in-vacuum binary PFL with 12\% solid angle coverage (NA=0.64). Ion size is limited by residual ion motion. \textbf{b.}  Depth of focus measurement for ion image. Ion image size (FWHM) is plotted as a function of focal position. Fitting the size to  $y_0 \sqrt{ 1+ (z/w_0)^2 }$ gives a depth of focus $2 w_0= 19.4\pm2.4 \:\mu$m and a minimum spot size $y_0=3.7\pm0.3 \:\mu$m. Uncertainty is dominated by pixel quantisation.}
\label{FigImgAndDOF}
\end{figure*}

Figure 2a shows an image of a single $^{174}$Yb$^+$ ion obtained from the fluorescence of the 369.5 nm transition. Images with full width at half maximum (FWHM) spatial sizes down to $3.7\pm0.3 \:\mu$m were obtained, limited by the residual ion motion. This spatial resolution is similar to existing trapped-ion QIP experiments and sufficient for scalable quantum state readout. The independently measured resolution of the PFL\cite{Streed-09} is substantially better and can be achieved by eliminating the residual ion motion. To estimate the alignment sensitivity of the PFL we measured the depth of focus and the field of view.  The depth of focus was determined by changing the position of the camera along the optical axis and inferring the equivalent change in the position of the ion. The resulting fit (Fig 2b.) of image size as a function of position gives a depth of focus of $19.4\pm2.4\:\mu$m and a minimum spot size of $3.7\pm0.3\:\mu$m. Likewise the field of view was inferred to be $140\pm20\:\mu$m by translating the RF needles $\pm50\:\mu$m from their center position and measuring the increase in ion spot size. In both cases (depth of focus and field of view) the imaging boundaries are defined as the range over which the observed ion spot area doubles. Given that currently used trap electrode feature sizes are \cite{Amini-10} on the order of tens of $\mu$m, the measured alignment tolerances indicate the viability of PFL arrays for use in massively parallel trapped-ion QIP.

To determine the fluorescence collection efficiency, we compared the experimentally observed detection rate with the expected ion scatter rate. Saturation of the ion florescence in the experiment produced a detection rate of $22.6 \pm 0.3 \times 10^3 \: \mbox{s}^{-1}$. The expected rate is calculated based on the combination of the maximum scatter rate of the ion, corrected for the camera's quantum efficiency, optical losses, and residual ion motion. Given the solid angle coverage of 12\% and ion transition natural linewidth of $\Gamma/2\pi=19.6$ MHz, we calculate a maximum saturated flux of $7.39\times10^6 \:\mbox{s}^{-1}$ at the PFL for an ion at rest.

We calibrated the quantum efficiency of our camera by illuminating the CCD sensor with an attenuated laser beam whose intensity, wavelength, and spatial profile closely matched that of the ion signal. The output from a diode laser was attenuated with a series of filters and then coupled into a single mode fiber. Light from the fiber tip was then imaged onto the camera with a spot size a few pixels wide, similar to that of the ion. Four optical attenuators of $3.2\pm0.1$, $43.2\pm0.1$, $27.7\pm0.1$, and $12.6\pm0.1$ dB were independently measured and combined to produce a total attenuation of $87\pm1$dB. To ensure that no light was leaking around the filters into the fiber from other sources, we measured the attenuation of pairs of these filters. Variations in these measurements constitute the dominant source of uncertainty in the total attenuation. A laser power of $30\pm1 \:\mu$W produced a background-corrected signal level on the camera of $33.0\pm0.3\times10^3 \:\mbox{s}^{-1}$. From this we calculate a quantum efficiency of $28\pm6\%$. Correcting for the camera's quantum efficiency as well as for optical losses from a 370 nm line filter (Thorlabs FB370, measured transmission $25\pm5\%$) and an uncoated vacuum window (calculated transmission 92\%) we infer a total photon flux at the PFL of $3.5\pm0.9\times10^5 \:\mbox{s}^{-1}$.

Residual ion motion reduced the scatter rate below that expected for an ion at rest. Stray electric fields in the trap chamber pushed the ion away from the node of the RF quadrupole field. The resulting micromotion broadened the linewidth and reduced the fluorescence of the ion\cite{Berkeland-98}. Since the RF drive frequency of 20 MHz and the transition's natural linewidth of $\Gamma/2\pi=19.6$ MHz are nearly the same, the micromotion has an effect similar to homogenous broadening, because of the unresolved motional sidebands and because the ion experiences a large fraction of the possible micromotion velocities during an excited state lifetime. The effects of the micromotion dominate the observed ion linewidth of $162\pm10$ MHz FWHM which shows the characteristic spectral ``scalloped'' shape\cite{Berkeland-98}. To this spectral feature we fit a modulation depth parameter $\beta=7.6\pm0.5$. The observed saturation intensity $I_{sat}=1.1\pm0.3\times10^3 \:\mbox{mW cm}^{-2}$ is $11\pm3$ times greater than that for an ion at rest, also consistent with a homogenous broadening model. The maximum scattering rate is thus reduced to $14.5\pm1.5\%$ of that for an ion at rest. Including these corrections we infer an ion fluorescence collection efficiency of $4.2\pm1.5\%$. Dividing this efficiency by the solid angle coverage gives a PFL diffraction efficiency of $35\pm13$\% for this lens, in agreement with our independently measured diffraction efficiency of $30\pm1\%$\cite{Streed-09}. Our ion fluorescence collection efficiency exceeds that of many recent trapped-ion QIP experiments \cite{Myerson-08, Olmschenk-10,VanDevender-10} and is approaching the proposed 5\% threshold for massively parallel implementations\cite{Steane-07}.

Imaging of trapped ions is extremely sensitive to stray light. Scattered light from nearby electrodes could couple into parasitic diffractive orders of the PFL and overlap with the ion image. This was not a significant issue in our system. The ion image contrast (ion signal rate to background rate) was $23\pm4$, even though our laser beam's $1/e^2$ diameter was only $2.5\times$ smaller than the needle separation. Eliminating the residual ion motion will increase the ion fluorescence while reducing the background rate, improving the image contrast to $\gtrsim 160$, comparable to the current state of the art for high-fidelity state readout\cite{Myerson-08}.

%%% Conclusion
In conclusion, we have imaged a trapped ion with a microfabricated optic for the first time. The ease of microfabricating large PFL arrays make them an attractive optical interconnect for massively parallel trapped-ion QIP\cite{Kielpinski-02, Amini-10}. The demonstrated collection efficiency and image contrast are competitive with other trapped-ion QIP experiments and suitable for large-scale QIP. Further improvements of the PFL to 28\% solid angle coverage \cite{Menon-06} and 80\% diffraction efficiency\cite{Cruz-Cabrera-07} would increase the collection efficiency to 22\%, more than double that recently reported with bulk optics\cite{Shu-10}. Light-induced charging can affect trapping of ions near a dielectric surface \cite{Harlander-10}, but the distance to the surface can be made as small as 80 $\mu$m, \cite{VanDevender-10} so that lenses only 500 $\mu$m in diameter give 28\% solid angle coverage. The field of view and the depth of focus are compatible with tolerances in current trapped-ion QIP microfabrication techniques\cite{Home-09,Amini-10}. Neutral\cite{Knoernschild-10} and solid state\cite{Hadden-10} QIP architectures also rely on interfacing with arrays of strongly divergent optical sources. Our approach can be readily extended to highly parallel optical coupling in these systems.

Support provided by the Australian Research Council under DP0773354 (DK), DP0877936 (ES, Australian Postdoctoral Fellowship), and FF0458313 (H. Wiseman, Federation Fellowship) as well as the US Air Force Office of Scientific Research (FA2386-09-1-4015). The phase Fresnel lens was fabricated by Margit Ferstl at the Heinrich-Hertz-Institut of the Fraunhofer-Institut f\"{u}r Nachrichtentechnik in Germany. We thank W. Campbell and C. Monroe for helpful discussion.

%\bibliography{EWSBib2010Oct13}

\end{document}